\documentclass[onecolumn]{aa}  
\usepackage{natbib,amsmath,amssymb,graphicx,txfonts}
\def\jetp{JETP}

\def\halpha{H$\alpha$}
\def\kms{km\,s$^{-1}$}

\begin{document}

   \title{
On the electron-ion temperature ratio established by collisionless shocks
}

\titlerunning{On the electron-ion temperature ratio produced by collisionless shocks}
   \subtitle{}

   \author{Jacco Vink
          \inst{1}
          \and
          Sjors Broersen\inst{2}
          \and
          Andrei Bykov\inst{3}
          \and
          Stefano Gabici\inst{4}
          }

   \institute{
Anton Pannekoek Institute/GRAPPA, University of Amsterdam, PO Box
94249, 1090 GE Amsterdam, The Netherlands
         \and
Anton Pannekoek Institute, University of Amsterdam, PO Box
94249, 1090 GE Amsterdam, The Netherlands
          \and
A.F.Ioffe Physical-Technical Institute, St. Petersburg 194021, also St.Petersburg State Politechnical University, Russia
          \and
APC, Univ Paris Diderot, CNRS/IN2P3, CEA/Irfu, Obs de Paris, Sorbonne Paris 
Cit\'e, France
             }

\offprints{J. Vink, \email{j.vink@uva.nl}}

   \date{}

\abstract{
Astrophysical shocks are often collisionless shocks, in which
the changes in plasma flow and temperatures across the shock
are established not through Coulomb interactions, but through
electric and magnetic fields. An open question about collisionless
shocks is whether electrons and ions each establish their own post-shock
temperature (non-equilibration of temperatures), or whether they
quickly equilibrate in the shock region.
Here we provide a simple, thermodynamic, relation for the minimum electron-ion temperature ratios that
should be expected as a function of Mach number. The basic assumption is that the enthalpy-flux of the
electrons is conserved separately, but that all particle species should undergo
the same density jump across the shock, in order for
the plasma to remain charge neutral. 
The only form of additional electron heating that we allow for is adiabatic heating,
caused by the compression of the electron gas.
These assumptions results in an 
analytic treatment of expected electron-ion temperature ratio that agrees with observations
of collisionless shocks: at low sonic Mach numbers, $M_\mathrm{s}\lesssim 2$,
the electron-ion temperature ratio is close to unity,
whereas for Mach numbers above $M_\mathrm{s}\approx 60$ the 
electron-ion temperature ratio asymptotically approaches a temperature
ratio of $T_\mathrm{e}/T_\mathrm{i}= m_\mathrm{e}/\langle m_\mathrm{i}\rangle$.
In the intermediate Mach number range the electron-ion temperature
ratio scales as $T_\mathrm{e}/T_\mathrm{i}\propto M_\mathrm{s}^{-2}$.
In addition, we calculate the electron-ion temperature ratios under
the assumption of adiabatic heating of the electrons only, which
results in a higher electron-ion temperature ratio, but preserves
the $T_\mathrm{e}/T_\mathrm{i}\propto M_\mathrm{s}^{-2}$ scaling. 
We also show that for magnetised shocks the electron-ion temperature ratio
approaches the asymptotic value $T_\mathrm{e}/T_\mathrm{i}= m_\mathrm{e}/\langle m_\mathrm{i}\rangle$
for lower magnetosonic Mach numbers ($M_\mathrm{ms}$), mainly because for a strongly
magnetised shock the sonic Mach number is larger than the magnetosonic Mach
number ($M_\mathrm{ms}\leq M_\mathrm{s}$).
The predicted scaling of the electron-ion temperature ratio is in  agreement with observational data for
magnetosonic  Mach numbers between 2 and 10,
but for supernova remnants the relation requires that the inferred Mach numbers for the observations
are overestimated, perhaps as a result of upstream heating in the
cosmic-ray precursor.
In addition to predicting a minimal electron-ion temperature ratio, 
we also heuristically incorporate ion-electron heat exchange at the shock,
quantified with a dimensionless parameter $\xi$, which is the fraction of the enthalpy-flux
difference between electrons and ions that is used for equilibrating the electron and ion temperatures.
Comparing the model to existing
observations in the solar system and supernova remnants suggests
that the data are best described by $\xi \gtrsim 5\%$, but also provides
a hint that
the Mach number of some supernova remnant shocks may have been overestimated;
perhaps as a result of heating in the cosmic-ray precursor.
}


\maketitle

\section{Introduction}
Most astrophysical shocks, whether in the solar system, in the interstellar 
medium, or very large scale shocks in clusters of galaxies, are so-called
collisionless shocks. The change in flow and plasma parameters
across these shocks are not established through particle-particle
collisions (Coulomb interactions), but through 
collective effects 
\citep[electric and magnetic field fluctuations, see the textbooks  devoted to collisionless shocks by][]{tidman71,balogh13}. 
One of the questions about collisionless shocks is whether or not different types
of particles will be shock-heated to the same temperature or not.
For a high Mach number shock in a single fluid approximation characterised by
an adiabatic index of $\gamma=5/3$ the post-shock temperature
should, according to the Rankine-Hugoniot jump conditions, be 
\begin{equation}
kT= \frac{2(\gamma-1)}{(\gamma+1)^2} \mu m_\mathrm{p}V_{\rm s}^2=\frac{3}{16}\mu m_\mathrm{p} V_\mathrm{s}^2,\label{eq:kTstd}
\end{equation}
with $\mu$ the average mass of the particles in units of the proton mass $m_\mathrm{p}$, and $V_\mathrm{s}$ the shock velocity. 
For plasmas with solar abundances $\mu\approx 0.6$.\footnote{
Here $\mu<1$ because electrons also have to be taken into account,
and they have a mass that is much lower than the protons.}
For collisionless shock heating, i.e. in absence of Coulomb equilibration, 
it is not clear whether a single temperature $kT$ characterises the temperature
of each particle species, since  this requires equilibration of
energies between the ions and the electrons.
It is not a priori known how this equilibration should
be established, since electron-ion collisions are by definition negligible in collisionless
shocks. Instead it is often assumed that downstream of high Mach number, collisionless shocks the expected  temperature
for each species $i$ is given by
\begin{equation}
kT_i= \frac{3}{16} m_\mathrm{i} V_{\rm s}^2,\label{eq:tstd}
\end{equation}
which implies that electrons have a temperature  a factor 1/1836 lower than the protons
\citep[e.g.][for review]{ghavamian13}.
Note that further downstream of the shock Coulomb collisions will tend to slowly
equilibrate the ion and electron temperatures. But plasmas in objects
like young supernova remnants (SNRs) are likely not to have reached full temperature
equilibrium throughout the entire shell \citep{itoh78,vink12,ghavamian13}.
The non-equilibration of ion and electron temperatures is not only important
for SNRs, but appears also to affect low Mach number shocks in the solar system
\citep[e.g.][]{schwartz88}, clusters of galaxies \citep{russell12}, and may
affect the observability of the yet to be detected warm hot intergalactic medium
\citep[WHIM,][]{bykov08c}, in which a fraction of 40-50\% of the  baryons in the
Universe reside.

Clear observational evidence for non-equilibration of electron and ion temperatures
comes from SNRs. Early evidence for non-equilibration
was provided by the fact that young SNRs seem to have lower {\em electron} temperatures
($kT_\mathrm{e} \lesssim 4$~keV) than they should have given that they have shock speeds
up to $\sim 5000$~\kms ($kT \sim 30$~keV). More direct proof that SNR shocks
heat the electrons to lower temperatures than the ions comes from 
measurements of the ion temperatures, obtained from thermal Doppler broadening of line emission in the optical 
\citep{rakowski03,ghavamian07},  UV
\citep{raymond95}, and X-rays \citep{vink03b,furuzawa09,broersen13}.
The most extensive results come from optical spectroscopy, 
using the broad-line component of \halpha\ emission,
which arises from charge exchange between neutral hydrogen penetrating
the shock and already shock-heated protons downstream of the shock.
The ratio between the narrow-line \halpha\ component,
caused by direct excitation, and the broad-line component is sensitive to 
the ratio of the electron to ion temperature \citep{ghavamian07,vanadelsberg08}. 
In some cases broad-line \halpha\ measurements can be supplemented
by X-ray measurements of the electron temperature \citep{rakowski03,helder11}.
These measurements suggest that
for shock velocities $\lesssim 400$~\kms\ the electron and proton
temperatures are more or less equal, whereas for higher velocities
the degree of equilibration appears to decrease roughly as
$T_{\rm e}/T_{\rm p}\propto V^{-2}$ \citep{ghavamian07}.

Observations of electron heating at the Earth bowshock shows that
electrons are in most cases heated to lower temperatures than the ions
\citep{schwartz88}, with a Mach number dependence
that suggests that  there is an inverse
correlation between Mach number and electron-ion temperature ratio \citep{ghavamian13}.
Finally, measurements of post-shock temperature profiles
in clusters of galaxies indicate that the electrons are heated at the shock
to similar temperatures as the ions \citep{markevitch06}, or they obtain slightly
lower temperatures than the ions \citep{russell12}.

In this paper we provide an alternative, simple  explanation for the  behaviour
of the electron-ion equilibration as a function of Mach number.
The approach does not rely on details of the shock
heating mechanism itself, but only on the thermodynamics of the shocks.
The only assumption that is made is that the density jump across the shock
is the same for all species, in order to maintain charge neutrality.
We show that this assumption does not support the
sometimes expressed idea that
non-equilibration implies that 
$T_\mathrm{e}/T_\mathrm{i}\approx m_\mathrm{e}/m_\mathrm{i}$.
Instead we show that 
for low Mach numbers $T_\mathrm{e}/T_\mathrm{i}\approx 1$,
and that only for high Mach numbers $T_\mathrm{e}/T_\mathrm{i}= m_\mathrm{e}/m_\mathrm{i}$. In between these extremes 
we obtain a relation $T_{\rm e}/T_{\rm p}\propto M^{-2}$, similar to the relation suggested
by \citet{ghavamian07}. 
A further modification to the model is made
by parameterising ion-electron heat exchange, heuristically
allowing for additional heat flow from the ions to the electrons.

\section{Derivation of a relation between electron and ion temperatures}
\label{sec:method}

The shock jump conditions for a plane parallel unmagnetised shock
are given by the well known Rankine-Hugoniot relations
\begin{align}
\rho_1v_1=& \rho_2v_2,\label{eq:masscons}\\
\left(P_1 + \rho_1v_1^2\right)=& \left(P_2 + \rho_2v_2^2\right),\label{eq:momcons}\\
\left(\frac{\gamma}{\gamma -1} P_1 + \frac{1}{2}\rho_1v_1^2\right)v_1=&
\left(\frac{\gamma}{\gamma -1} P_2 + \frac{1}{2}\rho_2v_2^2\right)v_2,
                         \label{eq:enthcons}
\end{align}
with the subscripts $1$ and $2$ indicating the quantities upstream and downstream of the
shock respectively, $v$ the plasma velocity in the frame comoving with the shock, $\rho=n \mu m_\mathrm{p}$ the density (and with $n$ the
particle density), and $P=n kT$ the gas pressure.
The solution to this equation is that the density ratio between pre-shock and post-shock
plasma is 
\begin{equation}
\chi\equiv \frac{\rho_2}{\rho_1}= \frac{v_1}{v_2}=\frac{(\gamma+1)M_\mathrm{s}^2}{(\gamma-1)M_\mathrm{s}^2+2},\label{eq:chi}
\end{equation}
and that the downstream temperature is
\begin{equation}
kT_2= \frac{1}{\chi}\left[\frac{1}{\gamma M_\mathrm{s}^2} + \left(1-\frac{1}{\chi}\right)\right]\mu m_\mathrm{p} v_1^2,
\end{equation}
with $\gamma=5/3$ the adiabatic index, and $M_\mathrm{s}\equiv v_1/c_\mathrm{s}=\sqrt{\rho v_1^2/(\gamma P_1)}$ the sonic shock Mach number. For $M_\mathrm{s}\rightarrow \infty$ this becomes Eq.~\ref{eq:kTstd}.
In Sect.~\ref{sec:magnetised} we also discuss the effects of magnetic fields, in which case the Alfv\'en Mach number should also be taken into account.

Note that these solutions do not take into account the effects of cosmic-ray acceleration. 
Cosmic-ray acceleration can be taken into account in the framework of the Rankine-Hugoniot relations \citep[e.g.][]{vink10a},
by allowing for compression of the inflowing plasma caused by interactions with cosmic rays in a so-called
cosmic-ray precursor. In that case the actual gas shock, called sub-shock, will have a lower Mach number
than the overal Mach number, as the inflowing plasma has already been heated by
adiabatic compression or even non-adiabatic heating processes,
operating in the shock precursor. 
Equations~\ref{eq:masscons}-\ref{eq:enthcons} should
then be applied to the sub-shock alone, and also the relations we derive below should be strictly
applied to the sub-shock, and not to the overall shock structure (sub-shock plus cosmic-ray precursor region).

Applying the full shock jump relations to each particle species separately would
lead to both separate temperatures and to separate shock compression ratios $\chi$.
Although separate temperatures are to be expected for collisionless shocks, separate  compression
ratios are very unlikely, as it would lead to charge separation, and hence large scale electric fields.
The length scale over which charge separation is dissolved is typically of the order
of the Debye length $l_\mathrm{D}\approx \sqrt{ kT/(4\pi n_\mathrm{e}e^2)}\approx 0.22 n_\mathrm{e}^{-1/2}(T/10^7\,\mathrm{K})^{1/2}$~km,
which indicates the distance over which the electric potential energy equals the kinetic energy of the charge particles
\citep[e.g.][]{zeldovich66,spitzer62}. The Debye length 
is much smaller than the typical length scale over which the shock is established, 
$l_\mathrm{sh}\approx c/\omega_\mathrm{pi}=\sqrt{c^2m_\mathrm{p}/(4\pi n_\mathrm{i}e^2)}\approx
228 n_\mathrm{i}^{-1/2}$~km \citep[e.g.][]{bale03}. 
The ratio of these two length scales is $l_\mathrm{D}/l_\mathrm{sh}=\sqrt{kT/(m_\mathrm{p}c^2)}$, which for
non-relativistic shocks is much lower than one.

The most natural outcome of a collisionless shock seems, therefore, that the overall
shock jump conditions should be applied to the plasma as a whole, leading to only one
value for the compression ratio, $\chi=\rho_2/\rho_1=
\rho_{\mathrm{e},2}/\rho_{\mathrm{e},1}=\rho_{\mathrm{i},2}/\rho_{\mathrm{i},1}$,
with the indices $\mathrm{e}$ and $\mathrm{i}$ indicating respectively the electron and ion contributions.
This implies that we apply only Eq.~\ref{eq:momcons} and Eq.~\ref{eq:enthcons} 
to the particle species separately, as Eq.~\ref{eq:masscons}  is already determined by
the full set of jump conditions applied to the plasma as a whole.
Of the two remaining equations,  it seems more natural
to concentrate on the enthalpy-flux conservation (Eq.~\ref{eq:enthcons})
 for each separate species, because electrons and ions can easily change directions,
 influencing the individual momenta of the particles (momentum being a vector) 
 whereas the energy exchange between species is absent or very slow in collisionless plasmas.
In other words, we do expect some exchange of momentum between electrons 
and ions (mediated by the electromagnetic field at the shock), whereas energy exchange between
electrons and ions is more difficult.
 This implies that the enthalpy flux of each particle species
 is conserved separately, but that not necessarily the momentum flux of each species is conserved separately.
 An analogy that comes to mind is a ball bouncing elastically off a wall. The lightest of the objects in that case, the ball,
 preserves its kinetic energy, but the momentum has clearly changed (the ball reversed its direction).
Considering only Eq.~\ref{eq:enthcons} for each separate species implies that 
the heating of electrons is simply caused by redistribution and thermalisation of
the energy of the electrons, perhaps through elastic scatterings in the
shock regions, followed by  electron-electron equilibration.

To calculate what the effects are of this ansatz, it is useful to introduce
the following relations between temperature and sonic Mach numbers:
\begin{equation}
\frac{kT_1}{\mu m_\mathrm{p}v_1^2}=\frac{P_1}{\rho_1v_1^2}=\frac{1}{\gamma M_\mathrm{s}^2}\ , \ 
\frac{kT_1}{\mu_\mathrm{e} m_\mathrm{p}v_1^2}=\left(\frac{\mu}{\mu_\mathrm{e}}\right)\frac{1}{\gamma M_\mathrm{s}^2}\ ,\  
\frac{kT_1}{\mu_\mathrm{i} m_\mathrm{p}v_1^2}=\left(\frac{\mu}{\mu_\mathrm{i}}\right)\frac{1}{\gamma M_\mathrm{s}^2},\label{eq:mach}
\end{equation}
where we have used the notation that $\mu$ is the average mass of all species in units of the proton mass,
whereas $\mu_\mathrm{e}$ ($=m_\mathrm{e}/m_\mathrm{p}\approx 1/1836$) is the electron mass and $\mu_{\mathrm{i}}$ the average ion mass (protons and other ions)
in units of the proton mass. For solar abundance plasmas $\mu_{\mathrm{i}}\approx 1.27$.


\subsection{The most extreme case of non-equilibration of electron and ion temperatures}
\label{sec:simplest}
We now proceed by considering the electron-enthalpy flux separately:
\begin{align}
\left(\frac{\gamma}{\gamma -1} P_{\mathrm{e},2} + \frac{1}{2}\rho_{\mathrm{e},2}v_2^2\right)v_2=&
\left(\frac{\gamma}{\gamma -1} P_{\mathrm{e},1} + \frac{1}{2}\rho_{\mathrm{e},1}v_1^2\right)v_1.\label{eq:electronenthalpy}
\end{align}
Note that we assume that the electrons and ions have equal bulk velocities.
The downstream electron temperature $kT_{\mathrm{e},2}$ is contained in the pressure term, 
$P_{\mathrm{e},2}=n_{\mathrm{e},2}kT_{\mathrm{e},2}$. With the help of Eq.~\ref{eq:mach}, and assuming that
the electrons and ions are equilibrated upstream of the shock
(i.e. $kT_{\mathrm{e},1}=kT_{\mathrm{i},1}=kT_1$),
we obtain
\begin{align}\label{eq:tel}
\frac{kT_{\mathrm{e},2}}{\mu_{\mathrm{e}}m_{\mathrm{p}}v_1^2}=&
\frac{kT_{\mathrm{e},1}}{\mu_{\mathrm{e}}m_{\mathrm{p}}v_1^2}
+
\frac{1}{2} \left(\frac{\gamma-1}{\gamma}\right) \left(1-\frac{1}{\chi^2}\right)
=
\frac{1}{\gamma M_\mathrm{s}^2}\left(\frac{\mu}{\mu_{\mathrm{e}}}\right)
+
\frac{1}{2}\left(\frac{\gamma-1}{\gamma}\right)\left(1-\frac{1}{\chi^2}\right),
\end{align}
which for $\gamma=5/3, M_\mathrm{s}\rightarrow \infty$ (implying $\chi=4$) gives
\begin{equation}
kT_{\mathrm{e},2}=\frac{1}{2}\left(\frac{\gamma-1}{\gamma}\right)\left(1-\frac{1}{\chi^2}\right)\mu_\mathrm{e}m_\mathrm{p}v_1^2=\frac{3}{16}\mu_\mathrm{e}m_{\rm p}v_1^2,\label{eq:el_temp}
\end{equation}
which is the expected temperature if there is 
no electron-ion equilibration established by the shock.

\begin{figure*}
\centerline{
\includegraphics[trim=60 60 80 80,clip=true,angle=0,width=0.7\textwidth]{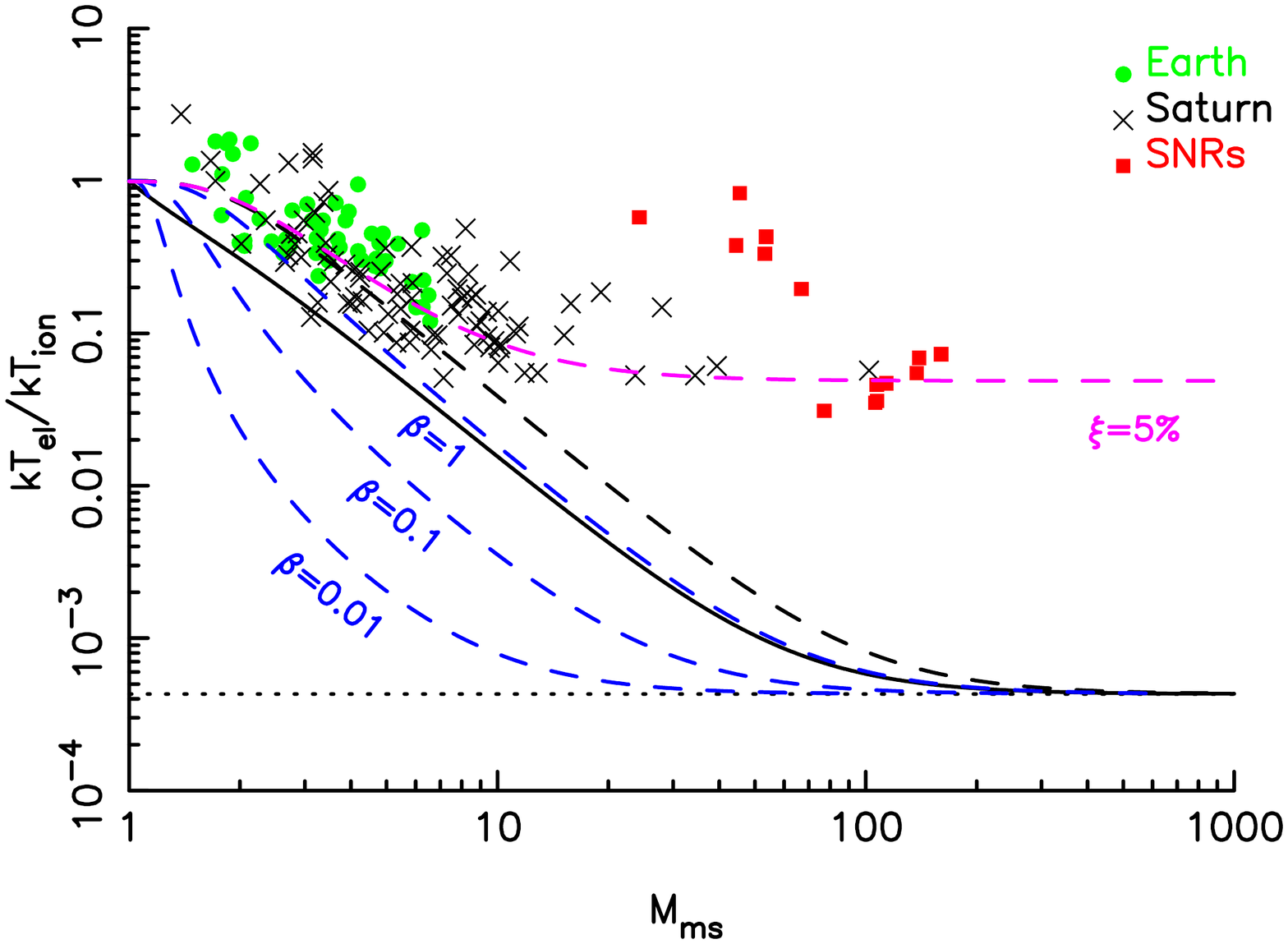}
}
\caption{
  The electron-ion  temperature ratio expected for
  minimum electron-heating (solid black line, Eq.~\ref{eq:Tratio}),
  for the case of adiabatic heating of the electrons (dashed black line, Eq.~\ref{eq:Tratio_ad}), and for magnetised plasmas with $\beta=0.01, 0.1, 1$ (blue dashed line, Eq~\ref{eq:Tratio_B}). In addition the expected ratio is shown if on top of various
  thermodynamic effects there is energy exchange between the electrons
  and ions at the level of 5\% (magenta dashed line, Eq.~\ref{eq:Tratio_xi}).
For the calculation here it is assumed that the plasma
consists of electrons, protons, and fully ionised helium, and the
ion temperature is the average temperature of abundance weighted temperature
of the protons and helium ions.
The data points are from the compilation by \citet{ghavamian13}, and represent measured
values of the electron-ion temperature ratio behind the
Earth bowshock \citep[green circles,original data from][]{schwartz88},
Saturn's bowshock  as measured by the Cassini spacecraft
\citep[X-shaped symbols,][]{masters13} and supernova remnants 
\citep[red, solid squares][]{vanadelsberg08}.
Note that for the supernova remnants the Mach numbers are not measured,
but instead the estimated shock velocity has been divided by assumed
interstellar sound speed of 11~\kms\ \citep{ghavamian13}.
}
\label{fig:equilibration}
\end{figure*}

In a similar way we can calculate the downstream ion temperature
by assuming ion-enthalpy flux conservation:
\begin{align}
\frac{kT_{\mathrm{i},2}}{\mu_{\mathrm{i}}m_{\mathrm{p}}v_1^2}=&
\frac{kT_{\mathrm{i},1}}{\mu_{\mathrm{i}}m_{\mathrm{p}}v_1^2}
+
\frac{1}{2}\left(\frac{\gamma-1}{\gamma}\right)\left(1-\frac{1}{\chi^2}\right)
=\frac{1}{\gamma M_\mathrm{s}^2}\left(\frac{\mu}{\mu_{\mathrm{i}}}\right)
+
\frac{1}{2}\left(\frac{\gamma-1}{\gamma}\right)\left(1-\frac{1}{\chi^2}\right).
\label{eq:tion}
\end{align}
Combining Eq.~\ref{eq:tel} with Eq.~\ref{eq:tion} we see that the
ratio of the electron temperature over ion temperature
is
\begin{align}\label{eq:Tratio}
\frac{T_{\mathrm{e},2}}{T_{\mathrm{i},2}}=
\left(\frac{\mu_\mathrm{e}}{\mu_\mathrm{i}}\right)
\frac{
2\left(\frac{\mu}{\mu_\mathrm{e}}\right)\chi^2
+M_\mathrm{s}^2(\gamma-1)\left(\chi^2-1\right)
}{
2\left(
\frac{\mu}{\mu_\mathrm{i}}
\right)\chi^2
 + M_\mathrm{s}^2(\gamma-1)
\left(\chi^2-1\right)}.
\end{align}
This relation is shown as a solid line in Fig.~\ref{fig:equilibration}.

For $M_\mathrm{s}\rightarrow \infty$ we find that  $kT_{\mathrm{e}}/kT_{\mathrm{i}}=\mu_\mathrm{e}/\mu_\mathrm{i}$, as usually
proposed for extreme non-equilibration, whereas for $M_\mathrm{s}\rightarrow 1, \chi\rightarrow 1$, we obtain
$kT_{\mathrm{e},2}/kT_{\mathrm{i},2}\rightarrow 1$. 
We see also that for 
\begin{equation}
\sqrt{\frac{2}{(\gamma-1)}\frac{\mu}{\mu_\mathrm{i}}\left(\frac{\chi^2}{\chi^2-1}\right)}< M_\mathrm{s} < 
\sqrt{\frac{2}{(\gamma-1)}\frac{\mu}{\mu_\mathrm{e}}\left(\frac{\chi^2}{\chi^2-1}\right)}\label{eq:mrange}
\end{equation}
(approximately $2<M_\mathrm{s}<60$)
 the term with $M_\mathrm{s}$ in the denominator of Eq.~\ref{eq:Tratio} is dominant, whereas $M$ is not yet  important for the nominator. As a result
the electron-ion temperature ratio in this  range of $M_\mathrm{s}$ scales as
$kT_{\mathrm{e},2}/kT_{\mathrm{i},2}\propto 1/M_\mathrm{s}^2$.

The result that the electron-electron-ion temperature ratios are close to one for low Mach numbers ($M_\mathrm{s}\approx 1$) and that
the ratio  decreases for  higher Mach numbers, can be made intuitively clear by realising that at relatively low
Mach numbers the enthalpy-flux is dominated by
the pre-shock thermal energy, which is similar for the electrons and the
ions (i.e. on average each particle has an  energy of $kT_1$  irrespective of its mass). 
For
high Mach numbers the enthalpy flux is dominated by the bulk
motion of the particles, $\frac{1}{2}nmv^3$. Hence, for large Mach numbers
the electron-enthalpy
flux is $m_\mathrm{e}/m_\mathrm{i}$ lower than the ion-enthalpy flux.

\subsection{Adiabatic heating of the electrons}
\label{sec:adiabatic}
A result of Eq.~\ref{eq:el_temp} is that for low
Mach numbers ($M_\mathrm{s} \lesssim 60$) the electron temperature is almost isothermal
across the shock. This would mean that the entropy of the electron gas
is decreasing, although the total entropy of the plasma is increasing.
The question is whether this is  physically a valid result or not.
In a closed system the entropy should aways be constant or increase,
but in this case the electrons are not an isolated system.
It is, therefore, more likely that the electrons are at least adiabatically
heated due to the compression caused by ions.
To take adiabatic electron heating into account
we first have to calculate the work done by the ions on the electrons
for adiabatic compression. 

We start with  modifying Eq.~\ref{eq:electronenthalpy} to include
an additional enthalpy-flux term, $q_\mathrm{e,ad}$, which is the excess
enthalpy flux needed to adiabatically compress the electrons:
\begin{align}
\left(\frac{\gamma}{\gamma -1} P_{\mathrm{e},2} + \frac{1}{2}\rho_{\mathrm{e},2}v_2^2\right)v_2=&
\left(\frac{\gamma}{\gamma -1} P_{\mathrm{e},1} + \frac{1}{2}\rho_{\mathrm{e},1}v_1^2\right)v_1 + q_\mathrm{e,ad}.\label{eq:electronenthalpy_ad}
\end{align}
For adiabatic compression the relation between pre- and post-shock 
electron pressures and temperatures is
\begin{equation}
P_{\mathrm{e},2}=P_{\mathrm{e},1}\chi^\gamma,\ T_\mathrm{e,2}=T_\mathrm{e,1}\chi^{\gamma-1}.
\label{eq:adiabat}
\end{equation}
If one  strictly wants to observe Eq.~\ref{eq:adiabat}  one obtains, from inserting Eq.~\ref{eq:adiabat} in Eq.~\ref{eq:electronenthalpy_ad}
\begin{equation}\label{eq:workwrong}
\frac{q_\mathrm{e,ad}}{\rho_{\mathrm{e},1}v_1^3}=
\left(\frac{\gamma}{\gamma-1}\right)\frac{kT_1}{\mu_\mathrm{e}m_\mathrm{p}v_1^2}
\left(\chi^{\gamma-1}-1\right) - \frac{1}{2}\left(1 - \frac{1}{\chi^2}\right).
\end{equation}

This expression for $q_\mathrm{e,ad}$ consists of two terms; the first
term ensures that the temperature will be proportional to $\chi^{\gamma-1}$, whereas the second term
suppresses the heating due to the kinetic energy flux of the electrons themselves.
The problem with this equation is exactly this second term, because  this term counteracts for  
high Mach numbers (i.e.  $M_\mathrm{s} \gtrsim 60$, as derived in Sect.~\ref{sec:simplest})
the heating of electrons from thermalising their own kinetic energy.
What is needed, therefore, is an expression for $q_\mathrm{e,ad}$ that enforces adiabatic heating
at low Mach numbers ($M_\mathrm{s}\lesssim 10$) through the  work done by the ions, 
whereas for $M_\mathrm{s} \gtrsim 60$ the electron kinetic energy flux is more than sufficient
to heat the electrons to temperatures beyond adiabatic compression.
These two conditions are met if one omits the second term on the right hand side of
Eq.~\ref{eq:workwrong}, i.e. the term that suppresses thermalisation of the electron kinetic energy.
Hence, the expression for the work done by the ions on the electrons becomes
\begin{equation}
\frac{q_\mathrm{e,ad}}{\rho_{\mathrm{e},1}v_1^3}=
\left(\frac{\gamma}{\gamma-1}\right)\frac{kT_1}{\mu_\mathrm{e}m_\mathrm{p}v_1^2}
\left(\chi^{\gamma-1}-1\right).
\label{eq:work}
\end{equation}
This expression and it is use in Eq.~\ref{eq:electronenthalpy_ad} seems a natural
choice, but is not completely from first principles. It assumes implicitly that the ions will
always first establish the shock, thereby compressing ad adiabatically heating the electrons, and that
subsequently the electron kinetic energy  is thermalised. In reality the electrons may be 
less passive when it comes to compressing the plasma, and part of the
energy needed for adiabatic heating is provided by the electrons themselves.
However, this is a concern only for intermediate Mach number shocks, since at Mach numbers below $M_\mathrm{s}\lesssim 10$
the electrons do not provide sufficient kinetic energy flux, whereas for  $M_\mathrm{s}\gtrsim 60$ the kinetic energy of the electrons
is more than enough to provide heating much beyond adiabatic heating, and the effects of introducing $q_\mathrm{e,ad}$ is neglible.

\begin{figure}
\centerline{
\includegraphics[trim=50 60 100 100,clip=true,width=0.6\textwidth]{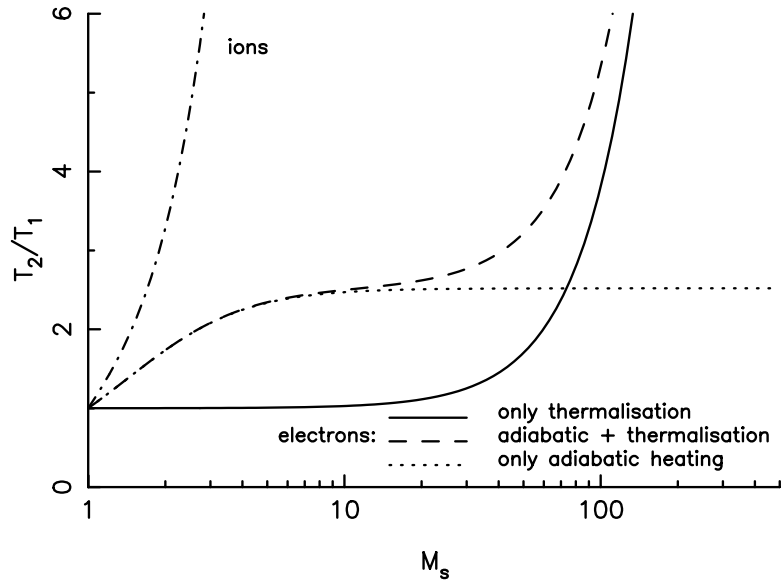}}
\caption{The ratio of the electron temperature upstream and downstream ($T_\mathrm{e,2}/T_\mathrm{i,2}$),
resulting from different assumptions. The dashed line corresponds to Eq.~\ref{eq:tel2} (adiabatic heating and thermalisation),
the solid line corresponds to Eq.~\ref{eq:tel}. For illustrative purposes we also
show the effect of adiabatic compression alone (dotted line) and the upstream and upstream temperature ratios
for ions (dot-dashed line).
\label{fig:adiabatic}
}
\end{figure}

Adopting Eq.~\ref{eq:work} for the enthalpy-flux exchange between ions and electrons
the expression for the downstream electron temperature becomes
\begin{align}\label{eq:tel2}
\frac{kT_{\mathrm{e},2}}{\mu_{\mathrm{e}}m_{\mathrm{p}}v_1^2}=&
\frac{kT_{\mathrm{e},1}}{\mu_{\mathrm{e}}m_{\mathrm{p}}v_1^2}\chi^{\gamma-1}
+
\frac{1}{2}\left(\frac{\gamma-1}{\gamma}\right)\left(1-\frac{1}{\chi^2}\right)
=
\frac{1}{\gamma M_\mathrm{s}^2}\left(\frac{\mu}{\mu_{\mathrm{e}}}\right)\chi^{\gamma-1}
+
\frac{1}{2}\left(\frac{\gamma-1}{\gamma}\right)\left(1-\frac{1}{\chi^2}\right).
\end{align}
Note that we  made use  of Eq.~\ref{eq:mach} in order to introduce the sonic Mach number.
One recognises in this equation again the two sources of heating, with the first term
being associated with adiabatic heating by the ions and the second term being associated with 
the thermalisation of the electron kinetic energy, which asymptotically approaches Eq.~\ref{eq:tstd}.

We illustrate the effects of adiabatic heating and thermalisation of the kinetic energy of the electrons in Fig.~\ref{fig:adiabatic}, 
which shows the ratio of downstream over upstream electron temperatures as given by Eq.~\ref{eq:tel2} and comparing
it to the result of Eq.~\ref{eq:tel} and  adiabatic heating only (i.e. if one would use $q_\mathrm{e,ad}$ as given in Eq.~\ref{eq:workwrong}). 
It shows that at low Mach numbers Eq.~\ref{eq:tel2} results in only adiabatic heating,
whereas for very high Mach number Eq.~\ref{eq:tel2} asymptotically approaches Eq.~\ref{eq:tel}.

The work done on the electrons will be at the cost of the heating of the
ions. We, therefore, have to subtract  $q_\mathrm{e,ad}$ (Eq.~\ref{eq:work}) from the ion-enthalpy flux:
\begin{align}\label{eq:ionflux_ad}
\left(\frac{\gamma}{\gamma -1} P_{\mathrm{i},2} + \frac{1}{2}\rho_{\mathrm{i},2}v_2^2\right)v_2=&
\left(\frac{\gamma}{\gamma -1} P_{\mathrm{i},1} + \frac{1}{2}\rho_{\mathrm{i},1}v_1^2\right)v_1 - q_\mathrm{e,ad}. 
\end{align}
Eq.~\ref{eq:ionflux_ad} and Eq.~\ref{eq:electronenthalpy_ad} added together just express that enthalpy-flux across the shock
is conserved (Eq.~\ref{eq:enthcons}).
This equation results in the following expression for the ion temperature:
\begin{align}
\frac{kT_{\mathrm{i},2}}{\mu_{\mathrm{i}}m_{\mathrm{p}}v_1^2}=&
\frac{kT_{\mathrm{i},1}}{\mu_{\mathrm{i}}m_{\mathrm{p}}v_1^2}\left[
1-\left(\frac{n_\mathrm{e,1}}{n_\mathrm{i,1}}\right)\left(\chi^{\gamma-1}-1\right)
\right]
+
\frac{1}{2}\left(\frac{\gamma-1}{\gamma}\right)\left(1-\frac{1}{\chi^2}\right)\\\nonumber
=&\frac{1}{\gamma M_\mathrm{s}^2}\left(\frac{\mu}{\mu_{\mathrm{i}}}\right)\left[
1-\left(\frac{n_\mathrm{e,1}}{n_\mathrm{i,1}}\right)\left(\chi^{\gamma-1}-1\right)
\right]
+
\frac{1}{2}\left(\frac{\gamma-1}{\gamma}\right)\left(1-\frac{1}{\chi^2}\right).
\end{align}
As a result the post-shock, electron-ion temperature ratio  will be
\begin{align}
\frac{T_{\mathrm{e},2}}{T_{\mathrm{i},2}}=&\left(\frac{\mu_\mathrm{e}}{\mu_\mathrm{i}}\right)
\frac{
2\left(\frac{\mu}{\mu_\mathrm{e}}\right)\chi^{\gamma+1} + M_\mathrm{s}^2\left(\gamma-1\right)\left(\chi^2-1\right)
}{
2\left(\frac{\mu}{\mu_\mathrm{i}}\right)\chi^2\left[
1-\left(\frac{n_\mathrm{e,1}}{n_\mathrm{i,1}}\right)\left(\chi^{\gamma-1}-1\right)
\right]+
 M_\mathrm{s}^2\left(\gamma-1\right)\left(\chi^2-1\right)
}.\label{eq:Tratio_ad}
\end{align}
Comparing this equation with Eq.~\ref{eq:Tratio} we see many similarities. The correctional factor in the denominator, given in square brackets,
is small as only a small fraction of the ion-enthalpy needs to be used to adiabatically heat the electrons. In the numerator we see that the factor
$\chi^2$ has been replaced by $\chi^{\gamma+1}$. For very high Mach numbers, roughly $M_\mathrm{s}>60$ as derived
in the previous section, the second term in the numerator becomes dominant and the electrons are heated beyond adiabatic heating.
However, the overall scaling of electron-ion temperature ratio is similar to Eq.~\ref{eq:Tratio}, namely $T_\mathrm{e,2}/T_\mathrm{i,2}\propto 1/M_\mathrm{s}^2$ for  $5\lesssim<M_\mathrm{s}\lesssim 60$, as can be seen in Fig.~\ref{fig:equilibration} (dashed black line). The temperature
ratio is, however, larger than given by  Eq.~\ref{eq:Tratio} (solid black line).

\subsection{The electron-ion temperature ratio in magnetised shocks}
\label{sec:magnetised}
In the previous subsections we have ignored the effects of magnetic fields. It is likely that the magnetic field and plasma waves play an important
role in the heating of the electrons and ions in collisionless shocks. However, here we focus on the importance of the shock thermodynamics
on the expected electron-ion temperature ratio. From a purely thermodynamic point of view the magnetic field
is important as it results in an additional pressure term, and the compression of the magnetic field requires additional
work to be done on the plasma.  The free energy-flux available for doing work
on the magnetised plasma will come mainly from the ion kinetic energy, as this is the largest reservoir of free energy. This suggests that
the downstream ion temperature will be lower for a given shock velocity if the shock is strongly magnetised.
As we will see below this is indeed the case as one compares magnetised shocks with unmagnetised shocks
for a given {\em sonic} Mach number, $M_\mathrm{s}$. However, for a magnetised shock the relevant Mach number is the {\em magnetosonic} Mach number,
$M_\mathrm{ms}$.
To illustrate the effects of the magnetic field strength on the electron-ion temperature ratio we calculate here
the thermodynamic effects of a magnetic field that is align strictly  perpendicular to the shock normal ($B_\perp=B$).

We parameterise 
the effects of the magnetic field using the plasma-beta parameter
\begin{equation}
\beta \equiv \frac{P_\mathrm{T,1}}{P_\mathrm{B,1}}=\frac{8\pi (n_\mathrm{i,1} + n_\mathrm{e,1})kT_1)}{B_1^2},\label{eq:plasmabeta}
\end{equation}
with $P_\mathrm{T,1}$ representing the total upstream, thermal pressure (electrons plus ions) and $P_\mathrm{B,1}$ the upstream magnetic pressure.
It is assumed here that upstream the electron and ion temperatures are equal ($kT_\mathrm{e,1}=kT_\mathrm{i,1}=kT_1$).
The ratio of the shock velocity to the Alfv\'en velocity, $V_\mathrm{A}=B_1/\sqrt{4\pi\rho_1}$, is the Alfv\'en Mach number, $M_\mathrm{A}=V_\mathrm{s}/V_\mathrm{A}$.
Note that the relation between Alfv\'en Mach number and sonic Mach number is given by the equation
\begin{equation}
M_\mathrm{s}^2=\frac{2}{\gamma \beta}M_\mathrm{A}^2.\label{eq:ms_vs_ma}
\end{equation}
These relations are useful, because the thermal pressure component of the downstream plasma is best parametrised by the sonic Mach number,
whereas the  shock compression ratio is mainly determined by the magnetosonic Mach number defined as
\begin{align}\label{eq:magnetosonic}
M_\mathrm{ms}= \sqrt{ \frac{1}{\frac{1}{M_\mathrm{s}^2} + \frac{1}{M_\mathrm{A}^2}}}=M_\mathrm{s}\sqrt{\frac{\gamma \beta}{\gamma \beta + 2}}.
\end{align}
The shock compression ratios for  perpendicular, magnetised shocks 
is given by \citet{tidman71} for $\gamma=5/3$, and can be expressed in the dimensionless numbers $\beta$ and $M_\mathrm{s}$ as
\begin{equation}
\chi= - \frac{5}{2}(1+ \beta) -\frac{5}{6}\beta  M_\mathrm{s}^2 + \sqrt{\frac{25}{4}\left(
1+\beta+\frac{1}{3}\beta M_\mathrm{s}^2
\right)^2+\frac{20}{3}\beta M_\mathrm{s}^2}.\label{eq:chimagnetised}
\end{equation}

There are two effects of magnetic fields on the electron-ion temperature ratio: 
1) for $\beta\ll 1$ the sonic Mach number is much larger
than the magnetosonic Mach number $M_\mathrm{s}\gg M_\mathrm{ms}$, as a consequence the electron-ion temperature ratio is expected to
be lower given Eq.~\ref{eq:Tratio} and/or Eq.~\ref{eq:Tratio_ad}; 
2) the kinetic energy of the ions will be partially used to do work on 
the magnetic field in order to compress it.
In analogy with Eq.~\ref{eq:work} we introduce here a parameter $q_\mathrm{B}$, which is the
 enthalpy flux necessary to compress the magnetic field:
\begin{align}
q_\mathrm{B}= &2P_\mathrm{B_\perp,2}v_2 - 2P_\mathrm{B_\perp,1}v_1=\frac{B_{\perp,2}^2}{4\pi}v_2 - \frac{B_{\perp,1}^2}{4\pi}v_1
= v_1 \frac{B_{\perp,1}^2}{4\pi}\left(\chi -1 \right),\\
\frac{q_\mathrm{B}}{\rho_1 v_1^3}=&\frac{B_{\perp,1}^2}{4\pi \rho_1 v_1^2}\left(\chi-1\right)=\frac{1}{M_\mathrm{A}^2}\left(\chi-1\right).
\end{align}

We  now proceed as  before by calculating the ion temperature from the enthalpy-flux equation for the ions only:
\begin{align}\label{eq:kTi_B}
\left(\frac{\gamma}{\gamma -1} P_{\mathrm{i},2} + \frac{1}{2}\rho_{\mathrm{i},2}v_2^2\right)v_2=&
\left(\frac{\gamma}{\gamma -1} P_{\mathrm{i},1} + \frac{1}{2}\rho_{\mathrm{i},1}v_1^2\right)v_1 - q_\mathrm{e,ad} -q_\mathrm{B},
\end{align}
which can be worked out to give the following downstream ion temperature
\begin{align}
\frac{kT_{\mathrm{i},2}}{\mu_{\mathrm{i}}m_{\mathrm{p}}v_1^2}=&
\frac{kT_{\mathrm{i},1}}{\mu_{\mathrm{i}}m_{\mathrm{p}}v_1^2}\left[
1-\left(\frac{n_\mathrm{e,1}}{n_\mathrm{i,1}}\right)\left(\frac{\mu_\mathrm{e}}{\mu_\mathrm{i}}\right)\left(\chi^{\gamma-1}-1\right)
\right]
+
\frac{1}{2}\left(\frac{\gamma-1}{\gamma}\right)\left(1-\frac{1}{\chi^2}\right)
-\frac{1}{M_\mathrm{A}^2}\left(\frac{\rho_1}{\rho_\mathrm{i,1}}\right)\left(\frac{\gamma-1}{\gamma}\right)\left(\chi-1\right).
\\\nonumber
=&\frac{1}{\gamma M_\mathrm{s}^2}\left(\frac{\mu}{\mu_{\mathrm{i}}}\right)\left[
1-\left(\frac{n_\mathrm{e,1}}{n_\mathrm{i,1}}\right)\left(\frac{\mu_\mathrm{e}}{\mu_\mathrm{i}}\right)\left(\chi^{\gamma-1}-1\right)
\right]
+
\frac{1}{2}\left(\frac{\gamma-1}{\gamma}\right) \left(1-\frac{1}{\chi^2}\right) 
- \frac{2}{\beta \gamma M_\mathrm{s}^2}\left(1 + \frac{n_\mathrm{e,1}}{n_\mathrm{i,1}}\frac{\mu_\mathrm{e}}{\mu_\mathrm{i}}\right)
\left(\frac{\gamma-1}{\gamma}\right)\left(\chi-1\right),
\end{align}
where we have made use of Eq.~\ref{eq:ms_vs_ma}, in order to make $M_\mathrm{s}$ and $\beta$ the primary shock parameters for magnetised shocks.
The factor in square brackets comes from the work done to
adiabatically heat the electrons (Sect.~\ref{sec:adiabatic}). 
If we only assume enthalpy-flux conservation for the electrons, this factor can be omitted. The factor $(\rho/\rho_\mathrm{i})$ is very close
to one.

We can now calculate the electron-ion temperature ratio by combining Eq.~\ref{eq:kTi_B} with Eq.~\ref{eq:tel2} , which results in
\begin{align}
\frac{T_{\mathrm{e},2}}{T_{\mathrm{i},2}}=&\left(\frac{\mu_\mathrm{e}}{\mu_\mathrm{i}}\right)
\frac{
2\left(\frac{\mu}{\mu_\mathrm{e}}\right)\chi^{\gamma+1} + M_\mathrm{s}^2\left(\gamma-1\right)\left(\chi^2-1\right)
}{
2\left(\frac{\mu}{\mu_\mathrm{i}}\right)\chi^2\left[
1-\left(\frac{n_\mathrm{e,1}}{n_\mathrm{i,1}}\right)\left(\frac{\mu_\mathrm{e}}{\mu_\mathrm{i}}\right)\left(\chi^{\gamma-1}-1\right)
\right]+
M_\mathrm{s}^2\left(\gamma-1\right)\left(\chi^2-1\right) -  \frac{4}{\beta \gamma} \left(1 + \frac{n_\mathrm{e,1}}{n_\mathrm{i,1}}\frac{\mu_\mathrm{e}}{\mu_\mathrm{i}}\right)
\left(\gamma-1 \right)\chi^2 \left(\chi -1 \right)
}.\label{eq:Tratio_B}
\end{align}
The dashed blue lines in Fig.~\ref{fig:equilibration} shows the derived electron-ion temperature ratio for various values of $\beta$.
Note that for $\beta=1$ the predicted ratio is close to an unmagnetised without
adiabatic heating.

For completeness' sake we also give here
the electrons the electron-ion temperature ratio for a magnetised shock
without adiabatic heating:
\begin{align}\label{eq:Tratio_B2}
\frac{T_{\mathrm{e},2}}{T_{\mathrm{i},2}}=&\left(\frac{\mu_\mathrm{e}}{\mu_\mathrm{i}}\right)
\frac{
2\left(\frac{\mu}{\mu_\mathrm{e}}\right)\chi^2 + M_\mathrm{s}^2\left(\gamma-1\right)\left(\chi^2-1\right)
}{
2\left(\frac{\mu}{\mu_\mathrm{i}}\right)
\chi^2+
M_\mathrm{s}^2\left(\gamma-1\right)\left(\chi^2-1\right) - 
\frac{4}{\beta \gamma}
\left(1 + \frac{n_\mathrm{e,1}}{n_\mathrm{i,1}}\frac{\mu_\mathrm{e}}{\mu_\mathrm{i}}\right)
\left(\gamma-1 \right)\chi^2\left(\chi -1 \right)
}.
\end{align}

\subsection{How much non-adiabatic heat exchange is there between ions and electrons?}
\label{sec:xi}

In the preceding subsections we have approached the expected
electron-ion temperature ratio from the point of view of the available
enthalpy of the electron and ion components separately, but allowing for
adiabatic heating of the electrons (Sect.~\ref{sec:adiabatic}) and the fact that
for magnetised plasma's work has to be done by mostly the ions to compress
the magnetic field (Sect.~\ref{sec:magnetised}).

However, collisionless shock heating is a complex process, and the microphysics
of the heating process of both the ions and electrons is not well understood.
In the Appendix (Sect.~\ref{sec:partialA}) we show how one can quantify the
additional thermal energy that the electrons pick-up from ions. This is done
by introducing an electron-ion heat exhange parameter $\xi$. A value
of $\xi=50$\% corresponds to a fully equilibrated plasma, whereas $\xi=0$ gives
the same result as Eq.~\ref{eq:Tratio_B}.

Fig.~\ref{fig:equilibration} shows the electron/ion temperature ratio for
$\xi=5$\%, which seems to approximately describe the levelling off of the
electron-ion temperature ratio seen in the Earth bow shock for magnetosonic
Mach numbers above $\sim 20$. However, more data are needed to see whether this
is a general trend, as not that much data points above $M_\mathrm{s}=10$ exist.
The supernova remnant shocks have generally higher Mach numbers, but for them
the actually Mach numbers are poorly constraint as we discuss below.

\begin{figure*}
\centering
\includegraphics[trim=50 60 115 100,clip=true,width=0.7\textwidth]{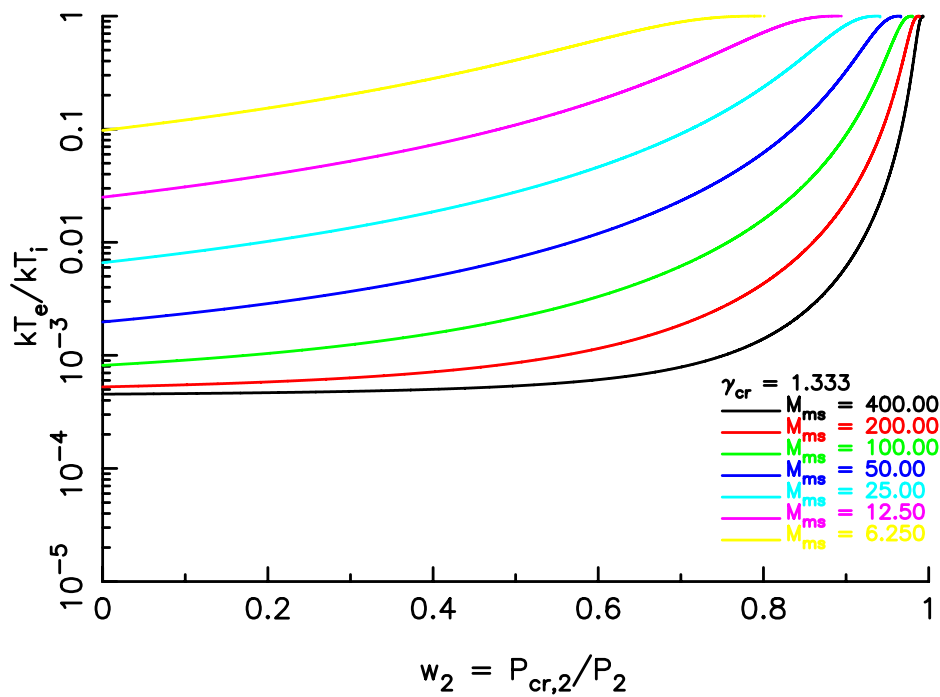}
\caption{
The electron-ion temperature ratio as a function of cosmic ray acceleration
efficiency. The efficiency is defined here as the fraction  $w_2$
of post-shock
cosmic-ray pressure (assumed here to be characterised by
an adiabatic index of $\gamma_{\rm cr}=1.5$) to the overall pressure. 
The cosmic-ray efficiency
is linked to the Mach number at the sub-shock, which has been affected by
the compression of the plasma in the cosmic-ray precursor. See
\citet{vink10a,vink14a} for details.
\label{fig:cr}
}
\end{figure*}

\section{Discussion}
The relation that we derived for the electron-ion temperature ratio is simple
and based on the assumption that all particle species observe the
same density jump, but that the electron and ion enthalpy fluxes are preserved
separately. The only additional heating of the electrons that we allow for (Sect.~\ref{sec:adiabatic})
is due to heating by adiabatic compression.
In this framework,
the potential role of the magnetic field (Sect.~\ref{sec:magnetised}) is passive:
magnetised shocks ($\beta<1)$ have a lower upstream electron/ion pressure 
for a given overall pressure.  Or to put it differently the sonic Mach number
is higher for a given magnetosonic Mach number, as compared to
plasmas with a larger plasma-beta. The electron/ion temperature ratio 
is in our explanation determined by the sonic Mach number and not by
the magnetosonic Mach number.

In addition, we introduced enthalpy-flow exchange between ions and electrons
using a dimensionless variable $\xi$, with (realistically) $0\% \leq \xi\leq 50\%$,
with $\xi=0$ corresponding to full non-equilibration. We stress that this is
an heuristic approach.

Comparing our relations with the measurements (Fig.~\ref{fig:equilibration})
shows that the electron-ion temperature ratio predicted
by the adiabatic heating case
(Eq~\ref{eq:Tratio_ad}) seems to describe the measurements 
best for Mach numbers between 1 and $\sim 10$, whereas the
the pure elastic scattering case (Eq.~\ref{eq:Tratio}) appears to
form a firm lower limit to the measured temperature ratios.
However, adiabatic heating of the electrons for magnetised plasma
with $\beta=1$ also seems to provide a lower limit to the measured
electron-ion temperature ratio. Magnetisation seems to be a more likely
explanation for the low electron/ion temperature ratios. 
For  Mach numbers $M_\mathrm{ms}\gtrsim 10$
the temperature ratio appears to asymptotically
reach a  value that is far above the  relations as predicted by
thermodynamic processes alone.

There are a few data points obtained from solar system shocks for
which the electrons appear hotter than the ions. These points cannot
be explained by our simple model. It may hint at possible heating of the electrons upstream of the sub-shock,
perhaps caused by ions reflected from the  subshock  \citep{cargill88}.
But it is not quite clear from the literature whether the measured temperature ratios
are not consistent with our relations given the unknown measurement errors,
and systematic uncertainties.

The measured SNR temperature ratios
 seem to behave  differently than the solar-system temperature ratios.
 As noted by  \citet{ghavamian13}, they do observe the relation
 $T_\mathrm{e}/T_\mathrm{i}\propto M_\mathrm{s}^{-2}$, but the data points
 are shifted to higher Mach numbers as compared to the solar system measurements.
But, as explained by \citet{ghavamian13},
for SNRs the actual
Mach number is not directly measured, but at best only the shock velocity
is measured. The Mach numbers are then inferred
by assuming that the local sound speed is 11~\kms. Assuming a higher
local, upstream sound speed would shift the SNR temperature
ratios closer to the solar-system values. This could either imply
locally larger temperatures for the interstellar medium, but it could
also mean that cosmic-ray precursors have pre-heated the plasma
before entering the shock, as already emphasised  by
 \citet{ghavamian13}. 

The effect of efficient cosmic-ray acceleration 
on the electron-ion temperature ratio can be estimated by
using the relations between 
post-shock, fractional cosmic ray pressure downstream of the shock, 
$w_2=P_\mathrm{cr}/P_\mathrm{tot}$, and adiabatic compression
in the cosmic-ray precursor, as derived in
\citet{vink10a} and \citet{vink14a}. 
The basic assumptions in these papers are that the cosmic-ray precursor induces a pre-compression upstream
of the shock $\chi_\mathrm{prec}$, resulting a in a lower Mach number at the actual gas shock
(sub-shock) to $M_\mathrm{s,sub}^2=M_\mathrm{s}^2\chi_\mathrm{prec}^{-\gamma}$.
The relation between fractional cosmic-ray pressure $w_2$, precursor compression,
and overall compression ratio, $\chi_\mathrm{tot}$ was derived by \citet{vink10a} to be
\begin{equation}
w_2=\frac{
\left(1-\chi_\mathrm{prec}^\gamma\right)+
\gamma M_\mathrm{s}^2\left(1-\frac{1}{\chi_\mathrm{prec}}\right)}{
1+\gamma M_\mathrm{s}^2\left(1-\frac{1}{\chi_\mathrm{tot}}\right)},
\end{equation}
Using these relations, we can calculate the electron-ion temperature ratio by
assuming that in the precursor ions and electrons are adiabatically heated 
to the same temperature and only at the sub-shock, with its reduced
Mach number, the electrons and ions are shock heated to the temperature
ratio given
by Eq.~\ref{eq:Tratio}.
The effect of the cosmic-ray  precursor on
the electron-ion temperature ratio is shown in Fig.~\ref{fig:cr}. 
Clearly the effects
are modest, except for very high cosmic-ray acceleration efficiencies ($w_2\gtrsim 0.5$). 
The effects could be larger if in the precursor also other,
non-adiabatic, heating processes play a role.There is observational
evidence, based on narrow line \halpha\ emission,
 that the upstream plasma of young SNRs is indeed hotter
than expected \citep{sollerman03}, perhaps as a result of heating in
the cosmic-ray precursor. {If the pre-heating of the neutral particles in the precursor is caused by
charge exchange of the neutral particles with the heated ions in the precursor \citep{raymond11,blasi12},
then the neutral particles provide a measure of the temperature in the precursor
at a distance $l\approx \tau_\mathrm{cx}v_1$ from the sub-shock (with $\tau_\mathrm{cx}$ the
average time between charge exchanges).
In contrast, the electron-ion temperature ratio is sensitive to the plasma temperature immediately
upstream of the subshock.
In fact, if by time we obtain sufficient
faith in the relations proposed in this paper, we may use the measured
electron-ion temperature ratio to infer the precursor temperature.
However, preferential heating of electrons upstream of the shock, as
perhaps indicated by some solar system shocks, may complicate the
use of the downstream electron-ion temperature ratio as derived here.
}

Apart from the effects of cosmic-ray acceleration on the electron-ion
temperature ratio, several other complications should be considered, which may,
under certain circumstances, affect the electron temperatures, or may affect the
interpretations of the
measurements. One non-trivial issue concerns the definition of 
the electron temperature itself. It presumes that the
electrons are thermalised, which for collisionless shocks may not always be the
case. For example, \citet{bykov99x} find that under certain circumstances
the energy distribution of the electrons is non-Maxwellian. 
On the other hand, if individual electrons are merely scattered elastically in the shock region,
the electron velocity distribution may be isotropised, but the electrons are in that
case almost mono-energetic, instead of having a Maxwellian distribution.
In either case, further
downstream of the shock the electrons may thermalise as a result
of electron-electron Coulomb interactions.
The timescale for electron-electron equilibration is
 relatively short ($\tau_\mathrm{ee}\approx 3.5 (T/10^6\,\mathrm{K})^{3/2}n_\mathrm{e}^{-1/2}$~months) compared to
the electron-ion equilibration time scale 
($\tau_\mathrm{ei}\approx 275 (T/10^6\,\mathrm{K})^{3/2}$~yr) \citep{spitzer62}.
For SNRs the measurable consequences may be limited, as the average electron
temperatures are typically measured further downstream of the shocks.
But for in situ measurements of temperatures downstream of
heliospheric shocks, the electron temperature measured may not correspond
to a well defined thermodynamic temperature. 
The shock jump conditions themselves
are valid for any energy distribution of the particles, although the relation
$P=nkT$ cannot be  blindly used, since a basic assumption of Eq.~\ref{eq:momcons} and Eq.~\ref{eq:enthcons} is that the pressure is isotropic. 
In essence  the pressure term in Eq.~\ref{eq:momcons}  refers to the
pressure along the shock normal, whereas in Eq.~\ref{eq:enthcons}
the pressure refers to the isotropic pressure, or internal energy. 
A last caveat to discuss is 
that our ansatz for calculating the expected electron-ion temperature
ratio assumes that the plasma upstream of the shock is fully ionised. Neutral
particles entering the shock will ionise further downstream of the shock and result
in a cold population of secondary electrons, 
which will slowly equilibrate with the primary electrons heated by the shock
itself \citep{itoh78}. This may affect the SNR measurements of the electron temperatures
in Fig.~\ref{fig:equilibration}, as most measurements are based on \halpha\ emission, 
which requires that the gas entering the shock is at least partially neutral.

As already discussed, measurements indicate that
$T_\mathrm{e}/T_\mathrm{i}\propto M_\mathrm{s}^{-2}$ over a limited range in $M_\mathrm{s}$,
with a lower limit to the temperature ratio at high Mach numbers, which indicates 
that there is at least some transfer of energy from the ions to the electrons.
We quantified this heat transfer
with the ion-electron cross-heating parameter $\xi$.
This is an heuristic approach that   hides potentially interesting microphysics.
For example, \citet{ghavamian07}, \citet{rakowski08} and \citet{laming14}
discuss the 
importance of lower hybrid-waves for heating of the electrons by ions.
In these models, which strictly applies to SNR shocks, the electron-heating
occurs in the cosmic-ray precursor region.
The idea that lower hybrid waves are needed
for giving a proportionality of $kT_\mathrm{e}/kT_\mathrm{i}\propto M^{-2}$ is
not necessarily true: as explained in Section~\ref{sec:method} and
as can be seen in Fig.~\ref{fig:equilibration},
this proportionality is a natural consequence of the Rankine-Hugoniot equations for Mach numbers
between $\sim 2$ and$ \sim 60$, with the upper limit depending on the
cross-heating parameter $\xi$. 
Other potential electron-heating mechanisms 
often rely on differences in flow speed between electrons and ions,
such as, for example, the Bunemann instability followed by  the ion-acoustic
instability \citep{cargill88},
which is caused by differences in flow speed between electrons and ions in the so-called
"foot" region of the subshock.

In our view the approach we have used shows that thermodynamics can explain part of the
electron-ion ratio that has been measured,  in particular the relation
$T_\mathrm{e}/T_\mathrm{i}\propto M_\mathrm{s}^{-2}$ for low Mach numbers. But 
for Mach numbers larger than ten the measurements indicate that additional electron heating
mechanisms may play a role, indicating a heat exchange between ions and electrons of
the order of 5-10\%. At low Mach numbers this additional heating may also play a role,
but is too small to leave a strong imprint on the electron-ion temperature ratio,
as can be seen in Fig.~\ref{fig:equilibration} (dashed line labeled $\xi=5$\%).

\section{Conclusion}

We have derived an equation that describes the electron-ion temperature
ratio, $T_\mathrm{e}/T_\mathrm{i}$ under the assumption that the
overall compression ratio follows the standard Rankine-Hugoniot relations
for the combination of electrons and ions, but that at the same time
the enthalpy flux of the electrons and ions can be treated separately
given the overall compression ratio. 
This assumption is valid if the electrons are heated
by  elastic scatterings in the shock region followed by electron-electron
thermalisation. In addition we derive expressions assuming
adiabatic heating of the electrons and we calculate the effects
of magnetisation.

All the relations we derive  give an electron-ion temperature ratio scaling
as $T_\mathrm{e}/T_\mathrm{i}\propto M_\mathrm{s}^{-2}$ for $2< M_\mathrm{s}< 60$, whereas
for higher Mach numbers  the ratio approaches asymptotically $T_\mathrm{e}/T_\mathrm{i} = m_\mathrm{e}/m_\mathrm{i}$.
Magnetisation produces the same asymptotic relation, but overall the electron-ion temperature
ratios are lower, mainly because for the temperatures of the ions and the electrons the sonic
Mach number, $M_\mathrm{s}$ is relevant, whereas the compression ratio is determined by
the magnetosonic Mach number $M_\mathrm{ms}$. For $\beta \ll 1$ $M_\mathrm{s}\gg M_\mathrm{ms}$.

In order to allow for heat exchange between electrons and ions we also
introduced the heat exchange factor $\xi$,  which is defined as
the fraction of the enthalpy-flux difference between ions and electrons that is used
to heat the electrons.
For increasing $\xi$ the electron-ion temperature levels off at 
increasingly higher values of  $T_\mathrm{e}/T_\mathrm{i}$. The available
data suggest that an appropriate value is at least $\xi=5$\%, whereas
$\xi \sim 50$\% corresponds to equal ion and electron temperatures.

\begin{acknowledgements}
SG and JV acknowledge the support from the Van Gogh programm for
exchange visits between Paris and Amsterdam.
SB was supported by an NWO Free Competition grant.
\end{acknowledgements}

\appendix


\section{Accounting for partial heat exchange}
\label{sec:partialA}
We can generalise our approach by allowing for some energy-flux
transfer from ions to electrons through non-elastic interactions
between electrons and ions. The interactions
do not have to be collisional, but could also
be mediated by electric and/or magnetic fields.

The available enthalpy flux from the
ions should then be corrected for the fact that some of the enthalpy flux
remains in the form of bulk kinetic energy downstream of the shock,
which  ensures that heat flows from the hottest component (ions)
to the coolest component (electrons).
This means that the available
ion-enthalpy flux that can be maximally transferred to the electrons is
\begin{align}
  q_\mathrm{i}\equiv&\left(\frac{\gamma}{\gamma -1} P_{\mathrm{i},1} + \frac{1}{2}\rho_{\mathrm{i},1}v_1^2\right)v_1 - \left(\frac{1}{2}\rho_{\mathrm{i},2}v_2^2\right)v_2 
  -q_\mathrm{e,ad}- q_\mathrm{B}\\\nonumber
 =&  
\frac{\gamma}{\gamma -1} P_{\mathrm{i},1} v_1  
+\frac{1}{2}\rho_{\mathrm{i},1}v_1^3\left(1 -\frac{1}{\chi^2}\right) 
-\left(\frac{\gamma}{\gamma-1}\right) P_\mathrm{e,1}\left(\chi^{\gamma-1}-1\right)v_1 -
\frac{B_{\perp,1}^2}{4\pi}\left(\chi - 1\right)v_1.
\label{eq:fion}
\end{align}

On the other hand it seems unphysical to assume that heat flows from one
component of the plasma to another if there is no difference in heat between
the two components.  We can make this explicit by 
specifying that the heat flow between ions and electrons must be proportional to
the difference between $q_\mathrm{i}$ and the equivalent quantity for the
electrons, $q_\mathrm{e}$. This difference is given by
\begin{align}
\Delta q= q_\mathrm{i}-q_\mathrm{e}=&
\left(\frac{\gamma}{\gamma-1}\right)P_{\mathrm{i},1}v_1 + 
\frac{1}{2}\rho_{\mathrm{i},1}v_1^3\left(1-\frac{1}{\chi^2}\right) - q_\mathrm{e,ad} - q_\mathrm{B} -
\left[\left(\frac{\gamma}{\gamma-1}\right)P_{\mathrm{e},1}v_1 + 
\frac{1}{2}\rho_{\mathrm{e},1}v_1^3\left(1-\frac{1}{\chi^2}\right) + q_\mathrm{e,ad}\right]\\\nonumber
=&
\left(\frac{\gamma}{\gamma-1}\right)
\left[n_{\mathrm{i},1}kT_{\mathrm{i},1}-n_{\mathrm{e},1}kT_{\mathrm{e},1}\left(2\chi^{\gamma-1}-1\right)\right]v_1
+
\frac{1}{2}m_\mathrm{p}
\left(n_{\mathrm{i},1}\mu_\mathrm{i}-n_{\mathrm{e},1}\mu_\mathrm{e}\right)
v_1^3\left(1-\frac{1}{\chi^2}\right) - v_1\frac{B_{\perp,1}^2}{4\pi}\left(\chi - 1\right).
\end{align}
We can now quantify the heat flux exchange between
electrons and ions in terms of this difference by
adding  $\xi \Delta q$ to Eq.~\ref{eq:electronenthalpy},  $\xi$ being  the fraction of the enthalpy-flux difference between
ions and electrons that is used to heat the electrons:
\begin{equation}
\left(\frac{\gamma}{\gamma -1} P_{\mathrm{e},2} + \frac{1}{2}\rho_{\mathrm{e,2}}v_2^2\right)v_2=
\left(\frac{\gamma}{\gamma -1} P_{\mathrm{e},1} + \frac{1}{2}\rho_{\mathrm{e},1}v_1^2\right)v_1 + q_\mathrm{e,ad} + \xi\Delta q,
\label{eq:electronenthalpy2}
\end{equation}
and subtracting the same term from the ion-enthalpy flux:
\begin{equation}
\left(\frac{\gamma}{\gamma -1} P_{\mathrm{i},2} + \frac{1}{2}\rho_{\mathrm{2}}v_2^2\right)v_2=
\left(\frac{\gamma}{\gamma -1} P_{\mathrm{i},1} + \frac{1}{2}\rho_{\mathrm{i},1}v_1^2\right)v_1  -q_\mathrm{B} -q_\mathrm{e,ad}- \xi\Delta q.
\label{eq:ionenthalpy2}
\end{equation}

Assuming that the upstream electron and ion temperatures are equal we obtain  the following expressions for the
downstream electron and ion temperatures:
\begin{align}
\frac{kT_\mathrm{e,2}}{\mu_\mathrm{e}m_\mathrm{p}v_1^2}=&
\frac{1}{\gamma M_\mathrm{s}^2}\left(\frac{\mu}{\mu_\mathrm{e}}\right)
	\left\{ 
			\chi^{\gamma-1} + \xi \left[
				\left(\frac{ n_\mathrm{i,1}}{n_\mathrm{e,1}}\right)-\left(2\chi^{\gamma-1}-1\right)
							\right]
	\right\}
	+
		\frac{1}{2}\left(\frac{\gamma-1}{\gamma}\right)\left(1-\frac{1}{\chi^2}\right)
		\left\{1 +\xi\left[\left(\frac{n_\mathrm{i,1}}{n_\mathrm{e,1}}\right)\left(\frac{\mu_\mathrm{i}}{\mu_\mathrm{e}}\right)-1\right]
		\right\} - \\\nonumber
                &			\xi \frac{2}{\beta \gamma M_\mathrm{s}^2}
                \left(1 + \frac{n_\mathrm{i,1}}{n_\mathrm{e,1}}\frac{\mu_\mathrm{i}}{\mu_\mathrm{e}}\right)
                \left(\frac{\gamma-1}{\gamma}\right)\left(\chi -1\right),
	\\
\frac{kT_\mathrm{i,2}}{\mu_\mathrm{i}m_\mathrm{p}v_1^2}=&
\frac{1}{\gamma M_\mathrm{s}^2}
\left(\frac{\mu}{\mu_\mathrm{i}}\right)
\left\{ 1- \left(\frac{n_\mathrm{e,1}}{n_\mathrm{i,1}}\right)
  \left(\chi^{\gamma-1}-1\right)-
  \xi \left[ 1 -
    \left(1 + \frac{n_\mathrm{e,1}}{n_\mathrm{i,1}}\frac{\mu_\mathrm{e}}{\mu_\mathrm{i}}\right)
    \left(2\chi^{\gamma-1}-1\right)
							\right]
	\right\}
	+
	\frac{1}{2}\left(1-\frac{1}{\chi^2}\right)
        \left(\frac{\gamma-1}{\gamma}\right)
		\left\{1 -\xi\left[1-\left(\frac{n_\mathrm{e,1}}{n_\mathrm{i,1}}\right)\left(\frac{\mu_\mathrm{e}}{\mu_\mathrm{i}}\right)\right]
		\right\} - \\\nonumber
                &			\left(1 - \xi\right) \frac{2}{\beta \gamma M_\mathrm{s}^2}
                \left(1 + \frac{n_\mathrm{e,1}}{n_\mathrm{i,1}}\frac{\mu_\mathrm{e}}{\mu_\mathrm{i}}\right)
                \left(\frac{\gamma-1}{\gamma}\right)
                \left(\chi -1\right).
\end{align}

Combining these two equations gives the electron-ion temperature ratio
\begin{align}\label{eq:Tratio_xi}
  \frac{T_{\mathrm{e},2}}{T_{\mathrm{i},2}}=&\left(\frac{\mu_\mathrm{e}}{\mu_\mathrm{i}}\right)
  \frac{
	\left(\frac{\mu}{\mu_\mathrm{e}}\right)\chi^2\left\{ 
			\chi^{\gamma-1} + \xi \left[
				\left(\frac{ n_\mathrm{i,1}}{n_\mathrm{e,1}}\right)-\left(2\chi^{\gamma-1}-1\right)
							\right]	\right\}
                        + M_\mathrm{s}^2\left(\gamma-1\right)\left(\chi^2-1\right)
                        \left\{1 +\xi\left[\left(\frac{n_\mathrm{i,1}}{n_\mathrm{e,1}}\right)\left(\frac{\mu_\mathrm{i}}{\mu_\mathrm{e}}\right)-1\right]\right\}
      -\xi \frac{4}{\beta \gamma }
                \left(1+\frac{n_\mathrm{i,1}}{n_\mathrm{e,1}}\frac{\mu_\mathrm{i}}{\mu_\mathrm{e}}\right)\left(\gamma-1\right)                  \chi^2\left(\chi -1\right)
  }{
\left(\frac{\mu}{\mu_\mathrm{i}}\right)
\chi^2\left\{ 1- \left(\frac{n_\mathrm{e,1}}{n_\mathrm{i,1}}\right)
  \left(\chi^{\gamma-1}-1\right)-
  \xi \left[ 1 -
    \left(\frac{ n_\mathrm{e,1}}{n_\mathrm{i,1}}\right)\left(2\chi^{\gamma-1}-1\right)
							\right]
	\right\}
	-
    	\left(1 - \xi\right) \frac{4}{\beta \gamma}
        \left(1+\frac{n_\mathrm{e,1}}{n_\mathrm{i,1}}\frac{\mu_\mathrm{e}}{\mu_\mathrm{i}}\right)
                \left(\gamma-1\right)        
        \chi^2\left(\chi-1\right)
}.
\end{align}
It can be easily verified that setting $\xi=0$ results in Eq.~\ref{eq:Tratio_B}.

\end{document}